\documentclass{moriond}

\usepackage{amsmath,amssymb}

\def\be{\begin{equation}}
\def\ee{\end{equation}}
\def\bea{\begin{eqnarray}}
\def\eea{\end{eqnarray}}

\newcommand{\Photo}{}

\begin{document}
\vspace*{4cm}
\title{Primordial black holes, a small review}

\author{A. Arbey}

\address{Universit\'e Claude Bernard Lyon 1, CNRS/IN2P3, \\
Institut de Physique des 2 Infinis de Lyon, UMR 5822, \\F-69622, Villeurbanne, France}

\maketitle\abstracts{
With the direct discovery of gravitational waves, black holes have regain interest in the recent years. In particular primordial black holes (PBHs), which originate from the very early Universe, may constitute (at least in part) dark matter. The possibility that dark matter is made of black holes is particularly appealing, and multi-messenger searches are important to probe this hypothesis. In this paper I will discuss the concept of primordial black holes, their origins, their characteristics and the current constraints. In addition I will explain that the study of black holes is of utmost interest since they may constitute portals to new physics and to quantum gravity.}

\vspace*{1.cm}

Black holes (BHs) are peculiar objects, in the sense that they question numerous fields beyond physics, such as philosophy, metaphysics, arts, etc. Their study is of capital importance for physics, as they represent at the same time limits to our spacetime and to our knowledge. In the following, we discuss the vast subject of primordial black holes (PBHs). More complete reviews can be found in Refs. \cite{Carr:2020gox,Auffinger:2022khh}.

\section{Primordial black holes are black holes!}

Primordial black holes represent a particular case of black holes, because \textit{primordial} means here that they found their origin in the primordial Universe. However following the \textit{no-hair conjecture}, a general relativity black hole is defined by only three parameters: mass, angular momentum (or spin) and electric charge. This conjecture relies on assumptions, such as the fact that black holes are stationary solutions of Einstein equations with flat space at infinity, which can therefore be questioned, but it is generally assumed that this conjecture holds and can even be a \textit{theorem}. As a consequence, it is impossible to differentiate black holes of different origins. Until now, all the observations of black holes are in agreement with the no-hair conjecture. In the following we will not discuss the case of electrically charged black holes since no observations of such objects have ever been made, and in this case one can assumes that black holes are determined by two parameters, mass $M$ and spin $J$, and they can be described in spherical coordinates $(t,r,\theta,\phi)$ by the Kerr metric:
\begin{equation}
 d\tau^2 = \big(dt - a \sin^2\theta \,d\phi\big)^2\, \frac{\Delta}{\Sigma} - \left(\frac{dr^2}{\Delta} + d\theta^2\right) \Sigma - \big((r^2+a^2)\,d\phi-a\,dt\big)^2 \, \frac{\sin^2\theta}{\Sigma}\,,
\end{equation}
where $a=J/M$, $\Sigma=r^2 + a^2 \cos^2\theta$, $\Delta = r^2 - R_s\,r + a^2$ and $R_s=2\,GM$ the Schwarzschild radius. One generally define $a^*=J/M^2$, and a requirement for the validity of this metric is that $0 \le a^* < 1$, $a^*=0$ corresponding to a standard non-rotating and spherically-symmetric Schwarzschild black hole. Black holes have a horizon which somehow corresponds to the spacetime surface on which the time-time metric term is zero. It is possible to find in-going solutions of the equations of motion but no out-going solutions, meaning that nothing can come out of black holes, justifying their name.

\textit{A priori} black holes can have any mass or spin, but only black holes with masses larger than a few solar masses have been observed, corresponding to stellar or supermassive black holes. Primordial black holes form in fact a broader class of black holes.

\section{Origins of primordial black holes}

The origin of stellar black holes is relatively well-known, and is related to the extremely high density which is reached during the core-collapse of a supernova, leading to an extreme curvature and the formation of a horizon. The idea behind the formation of primordial black holes is relatively similar: in the early Universe the energy density can be very large, and an over-density may lead to a curvature large enough to form a black hole. The existence of such an over-density is very model-dependent, and is generally considered to be related to phase transitions, formation and collapse of domain walls, inflation, or unknown mechanisms.

Let us have a closer look at the standard cosmological model: the Universe is in expansion with a scale factor $a(t)$, which is determined by the Friedmann-Lemaître equation:
\begin{equation}
H^2 = \left(\frac{\dot{a}}{a}\right)^2 = \frac{8\pi G}{3} \rho_{\rm tot}\,,
\end{equation}
where $H$ is the Hubble parameter and $\rho_{\rm tot}$ is the total energy density in the Universe at time (including in this case a curvature density related to the global geometry of the Universe). The basic assumption to solve this equation is that at $t=0$, $a(0)=0$, meaning a infinite density and leading to many philosophical and physical questions. Since the lengths are dilated by the scale factor $a(t)$, $H(t)$ is effectively the expansion rate, and $d_H = c / H$, which is called the Hubble distance, can be seen as the maximal size of causally related regions. With these ideas, one considers that the maximal size of a black hole is related to this Hubble distance. Assuming that a black hole has a horizon radius close to the Schwarzschild radius, assuming the cosmological standard model in which the Universe expansion is originally driven by radiation, one shows that the maximal mass of a PBH is:
\begin{equation}
M_{\rm PBH} \lesssim 10^{38}\;{\rm g}\;\times t(s)   \,, 
\end{equation}
where $t(s)$ is the cosmological time of PBH formation in seconds. This gives:
\begin{itemize}
 \item $M \sim 10^{-5}$ g at Planck time $t\sim10^{-43}$ s,
 \item $M \lesssim 10^{15}$ g for $t\sim10^{-23}$ s,
 \item $M \lesssim10^5\;M_\odot$ for $t\sim1$ s, and such PBHs can be seeds of supermassive black holes.
\end{itemize}
The main result is that primordial black holes can have any mass, provided one can find a mechanism to generate large over-densities in the very early Universe. For this reason PBHs are the most general cases of black holes. In addition, because of this, PBHs are often considered to be candidates for dark matter \cite{Arbey:2021gdg,Green:2020jor}.

Concerning the spin of black holes, it is again very model-dependent. Following the standard case where radiation dominates the expansion, spins are generally small \cite{DeLuca:2019buf}. Nevertheless, if many black holes are formed in the very early Universe, they will behave as a dark matter component which drives the expansion, therefore changing the expansion rate and generating a \textit{transient matter phase}. The difference between radiation and matter behaviours can lead to different formation mechanisms, which can generate high spins \cite{Harada:2017fjm}.

In any case, with a model-independent perspective, the mass and spin distributions are largely unknown, meaning that any mass or spin is possible, and some papers have even emitted the hypothesis that Planet 9 may be a black hole of the size of a tennis ball \cite{Scholtz:2019csj,Arbey:2020urq}.

\section{Quantum black holes}

A black hole of about $10^{15}\,$g is approximately of the same size as a nucleus. It is therefore easy to assume that quantum effects may affect such a black hole. Since no quantum gravity theory currently exists, semi-quantum approaches have been developed, in particular by S. Hawking~\cite{Hawking:1975vcx}, and primordial black holes are ideal objects to test such approaches \cite{Carr:1974nx}. The basic idea is that vacuum quantum fluctuations may generate pairs of particles, one of the particles falling into the horizon whereas the other one escapes. This process is relatively similar to black-body radiation, with the difference that the emitted particles have to escape the strong gravitational curvature generated by the black hole. For this reason, one speaks of \textit{grey-body radiation} and the rate of emission of particles $i$ at energy $E$ by a BH of mass $M$ and spin parameter $a^*$ reads
\begin{equation}
Q_i = \dfrac{{\rm d}^2 N_i}{{\rm d}t\,{\rm d}E} = \frac{1}{2\pi} \sum_{\rm dof.} \dfrac{\Gamma_i(M,E,a^*)}{e^{E/T(M,a^*)} - (-1)^{2s_i}} \,,
\end{equation}
which is similar to the Planck's radiation law. $\Gamma_i$ is the grey-body factor, acting as an absorption coefficient, $s_i$ the spin of the emitted particles and $T$ an effective temperature given by
\begin{equation}
T(M,a^*) = \frac{1}{4\pi M}\left( \frac{\sqrt{1 - (a^*)^2}}{1 + \sqrt{1 - (a^*)^2}} \right)\,.
\end{equation}

\begin{figure}[t]
\centerline{\includegraphics[height = 9cm]{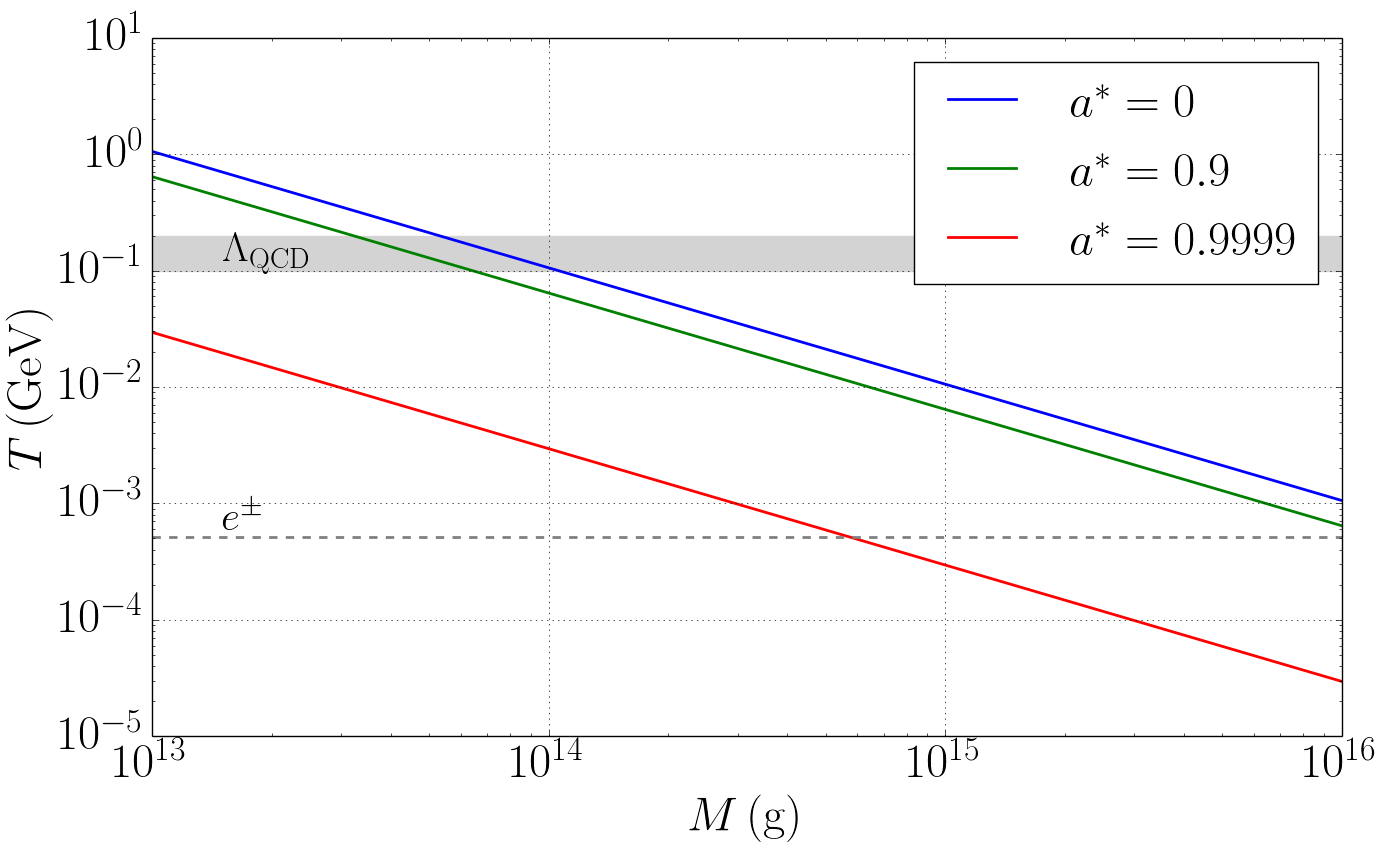}}
\caption{Black hole temperature as a function of the mass, for different values of the spin. For comparison, the electron mass and QCD scales are also shown.\label{fig:temp}}
\end{figure}

Figure~\ref{fig:temp} shows the effect of mass and spin on the temperature. The higher the mass, the smaller the temperature, resulting in reduced emission rates.

In practice, all types of particles can be emitted provided the temperature is high enough. Computing the emission spectra of particles is a difficult problem, and we have written the public code~~{\tt BlackHawk} \cite{Arbey:2019mbc,Arbey:2021mbl}~~to compute the primary and secondary spectra of Hawking radiation, which is used to generate plots shown in the following. 

\begin{figure}[t]
\centerline{\includegraphics[height = 11.cm]{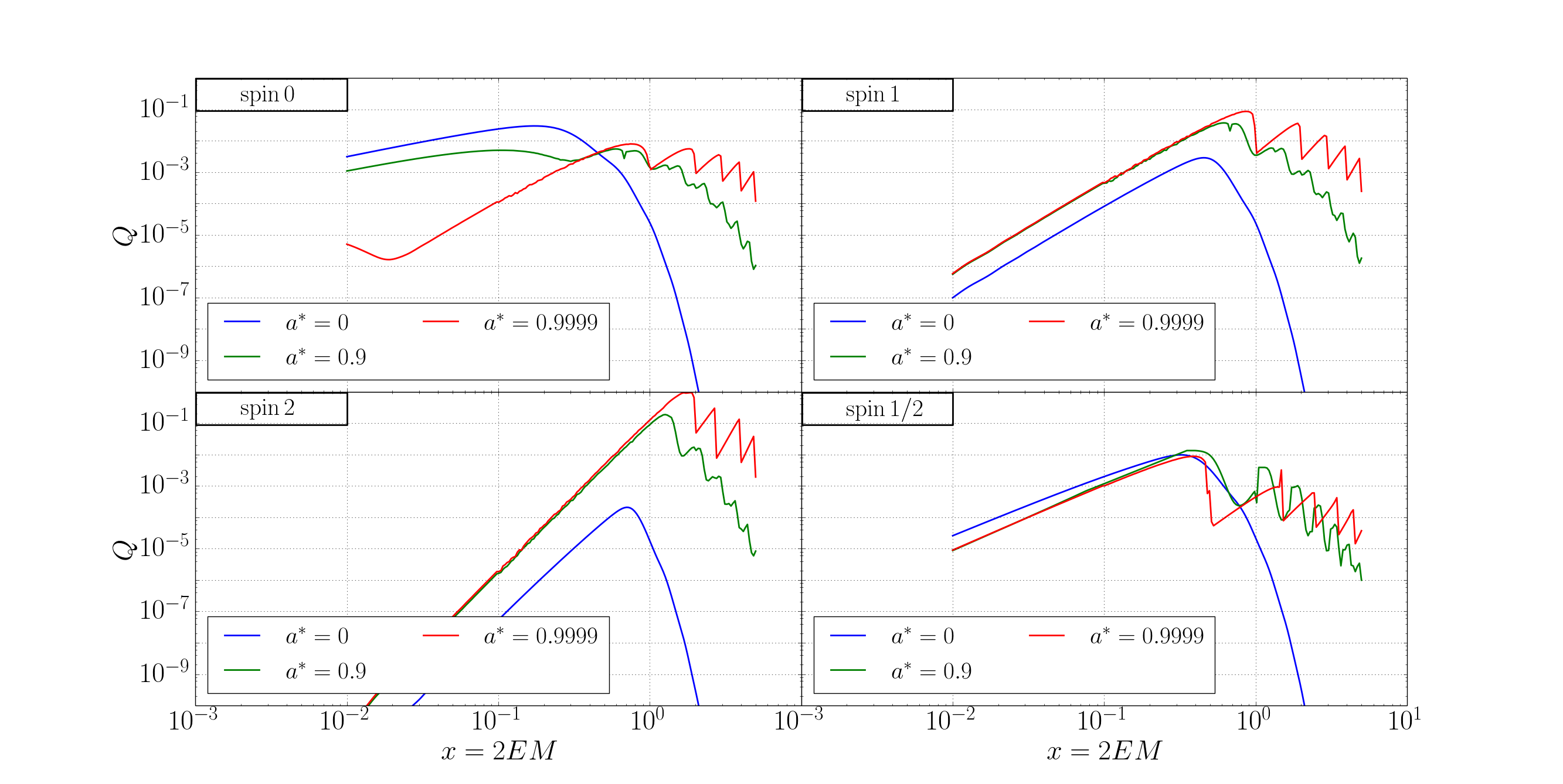}}
\caption{Emission rates of spin 0, 1, 2 and 1/2 particles of energy $E$ as functions of $2EM$, where $M$ is the black hole mass, for different spin values.\label{fig:rates}}
\end{figure}

In Fig.~\ref{fig:rates} we show the emission rates of black holes for the different types of particles, and the effect of the BH spin. In particular, the BH spin increases the emission rates for non-zero spin particles. It is important noticing that if gravitons exist, they can be emitted by BHs via Hawking radiation and detected as gravitational waves. 

\begin{figure}[t]
\centerline{\includegraphics[height = 7.5cm]{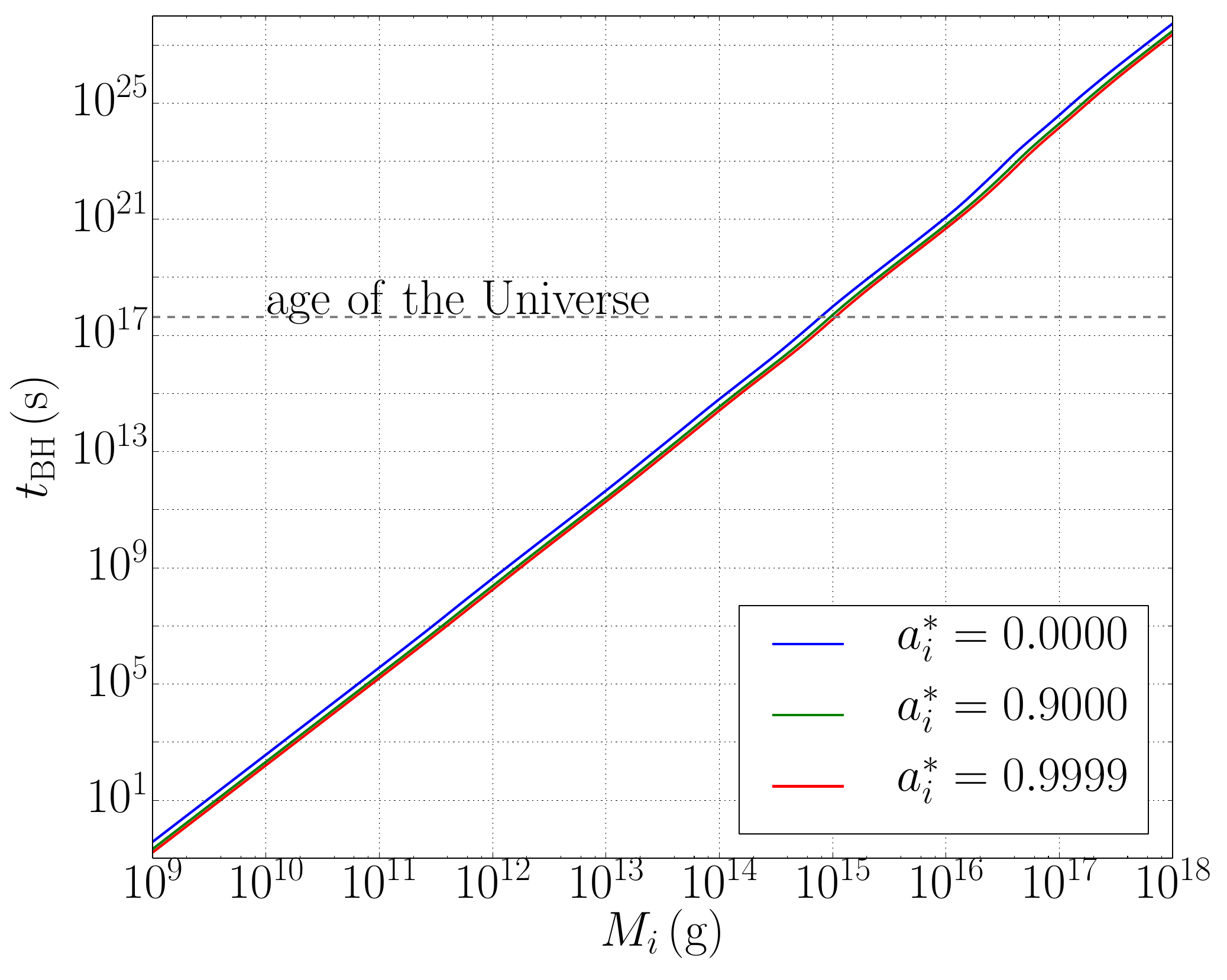}~\includegraphics[height = 7.5cm]{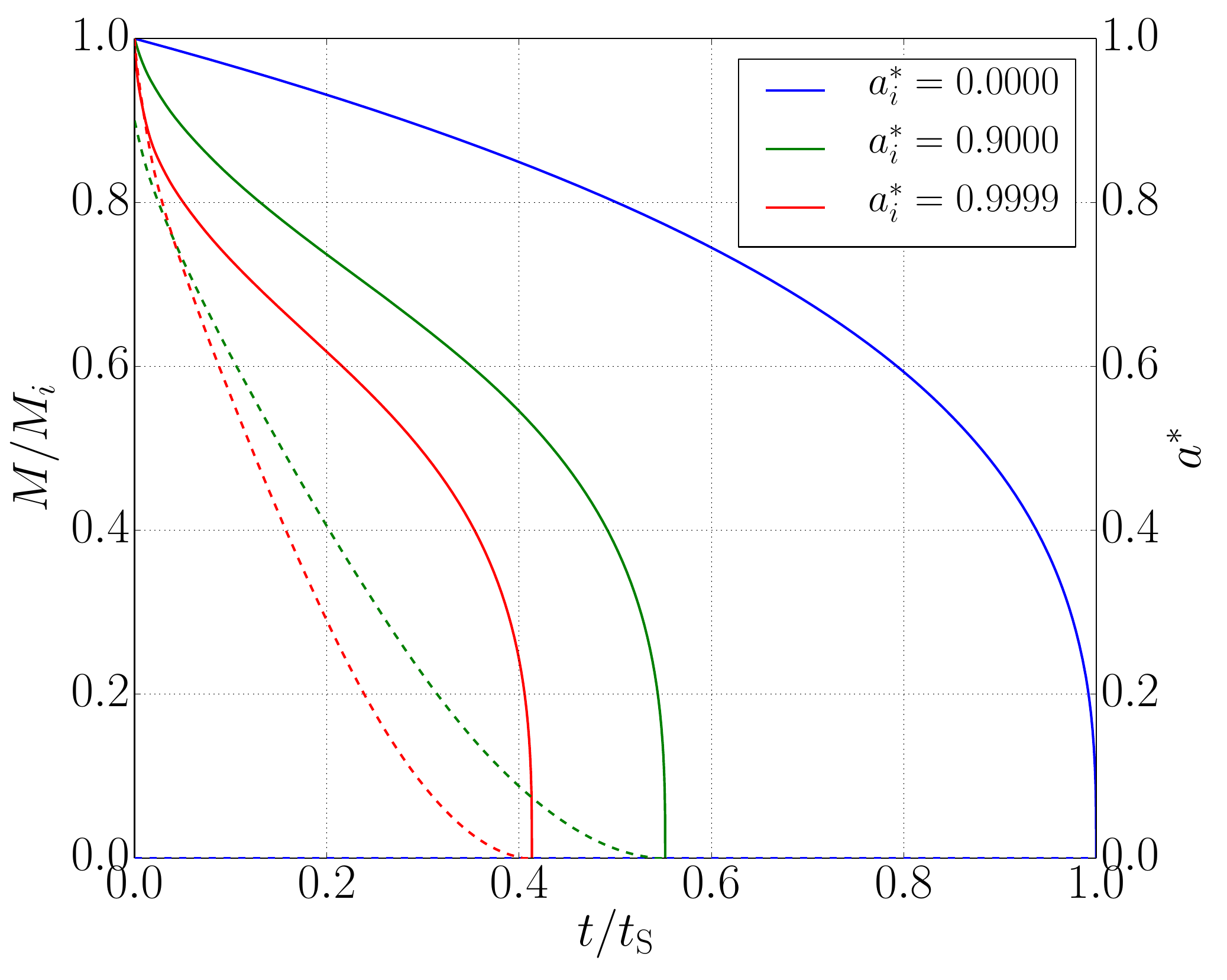}}
\caption{(Left) Black hole lifetime as a function of its mass, for different values of the spin. For comparison the age of the Universe is shown. (Right) Evolution of the mass (plain lines, scale on the left) and spin (dashed lines, scales on the right) as a function of time. The mass is normalised to the initial BH mass and the time to the same-mass Schwarzschild lifetime. From Ref. \protect\cite{Arbey:2019jmj}.\label{fig:evolution}}
\end{figure}

Since black holes are losing energy via particle emission, their mass and spin decrease. This is shown in Fig.~\ref{fig:evolution}. It reveals that BHs of about $10^{15}\,$g have lifetime similar to the age of the Universe, such that heavier BHs can be considered as stable, while lighter ones are \textit{evaporating} and may have already completely vanished since their creation time. BHs with masses close to $10^{15}\,$g can still exist and constitute dark matter, and their emission of particles could be detected by astroparticle experiments.

\section{Constraints on primordial black holes}

Constraints on light and on heavy PBHs are obtained differently.
For light PBHs, one searches for the particles emitted by Hawking radiation, using the results of astroparticle experiments. This can allow us to set constraints for BH masses below $\sim 10^{16}\,$g.
For heavy black holes, since the emission rate of Hawking radiation is too small, only gravitational effects can be used. In particular, the main observations are based on gravitational lensing, emission of gravitational waves by binary BHs, dynamical effects, large scales structures, accretion effects and distortions in the cosmic microwave background.
Up-to-now we have no evidence for the existence of \textit{primordial} black holes, but at least gravitational waves are proofs for the existence of black holes.

As a general fact, PBHs behave as dark matter, and the observed cosmological dark matter density can be considered as a maximum value for the PBH mass density.
On the other hand, since the initial PBH mass and spin distributions are unknown, one generally presents the constraints for Schwarzschild BHs (without spin) assuming a monochromatic mass distribution~\footnote{
An analysis of the effects of PBH mass distributions and of the spins can be found for example in Ref. \cite{Arbey:2019vqx}.} and considering that PBHs follow the same density distribution as dark matter. 

\begin{figure}[p]
\centerline{\includegraphics[height = 10.5cm]{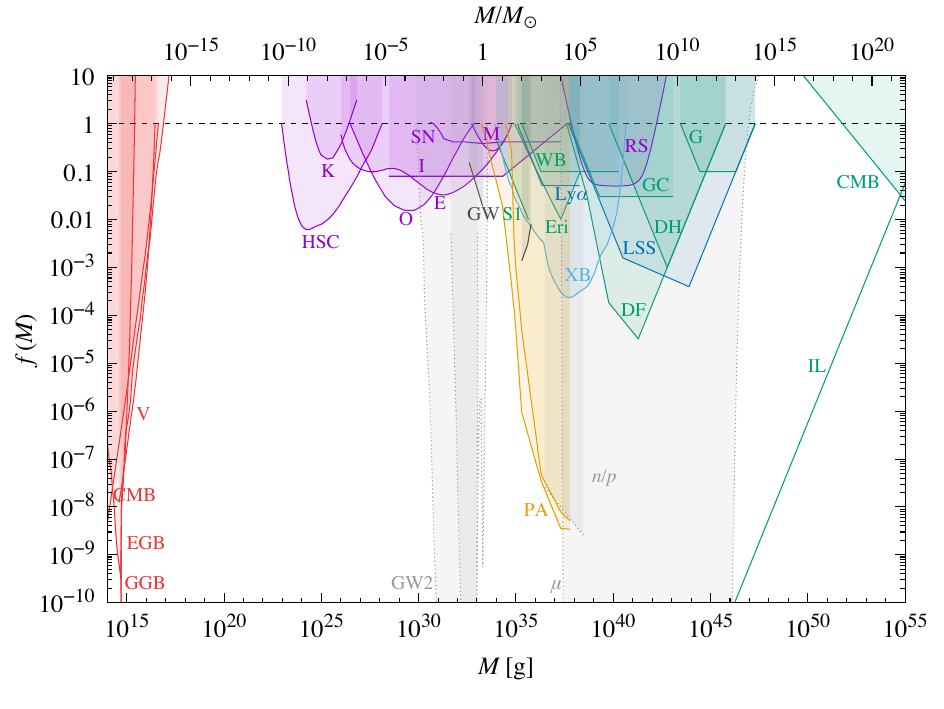}}
\caption{Fraction of PBHs to dark matter as a function of the monochromatic mass of PBHs. The coloured regions are excluded by observations. On the left side of the plot lighter PBHs vanished and do not exist anymore, and on the right side heavier PBHs are theoretically excluded. From Ref. \protect\cite{Carr:2020gox}.\label{fig:constraints}}
\centerline{\includegraphics[height = 10.5cm]{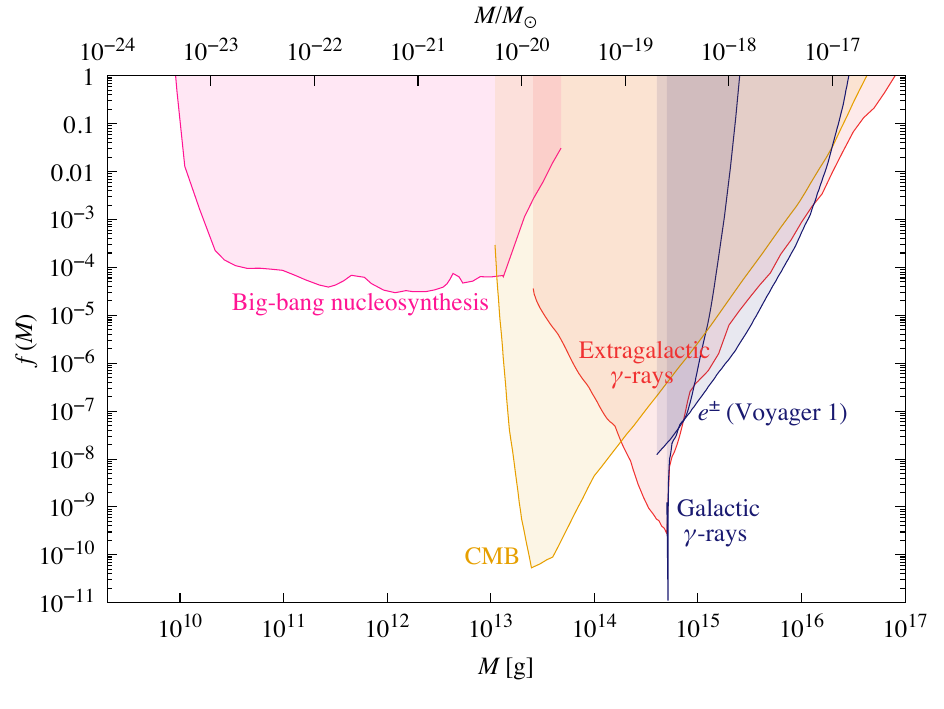}}
\caption{Fraction of PBHs to dark matter as a function of the monochromatic mass of PBHs, for light PBHs which have evaporated in the early Universe. The coloured regions are excluded by cosmological observations. From Ref. \protect\cite{Carr:2020gox}.\label{fig:constraints_evap}}
\end{figure}

Recent observational constraints are shown in Fig. \ref{fig:constraints}. One notices that there exist two windows in which PBHs can represent 100\% of dark matter, at low mass of about $10^{20}\,$g and at large masses of $10^{15}\,M_\odot$. For the other regions, PBHs can only represent a small fraction of dark matter, apart for PBHs with masses below $10^{15}\,$g which are completely excluded since they should not exist anymore because they vanished in the early Universe.

Using other cosmological observables it is possible to set constraints on the presence of very light PBHs in the early Universe, even if they already vanished. In particular, the evaporation of PBHs injects many particles in the early Universe, which can alter Big-Bang nucleosynthesis, recombination, or generate high energy particles which can be detected in current astroparticle experiments. Figure \ref{fig:constraints_evap} summarises the obtained constraints.

\begin{figure}[t]
\centering{\includegraphics[height = 9cm]{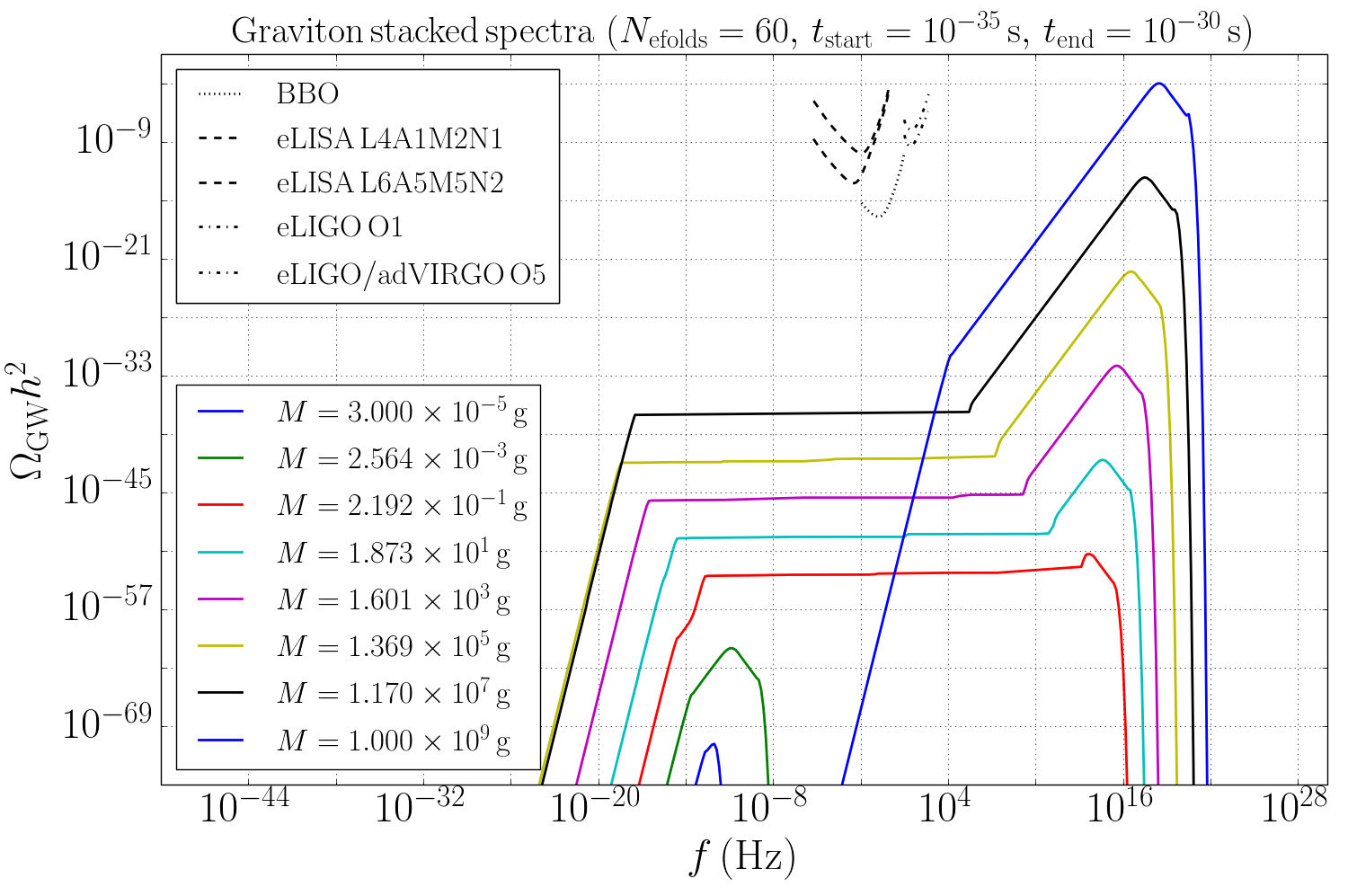}}
\caption{Energy density of the gravitational waves/gravitons emitted by very light PBH Hawking radiation as a function of their frequency, for different PBH masses. A standard inflation model with 60 e-folds is considered. The dashed lines corresponds to the reach of current and future gravitational wave experiments.\label{fig:graviton}}
\end{figure}

A way to set constraints at even earlier cosmological times would be to use the fact that very light black holes emit particles with a high luminosity. Unfortunately all these particles would be absorbed in the cosmological plasma. A way out would be to look for gravitons emitted by Hawking radiation. Figure \ref{fig:graviton} illustrates this idea: extremely light primordial black holes vanishing at the time of inflation can emit very high energy gravitons with an energy density similar to the sensitivity of the current and future gravitational wave experiments. Unfortunately, the very high frequency of the corresponding gravitational waves would require very small gravitational wave detectors very far from our current technology.

In order to interpret the obtained constraints, it is very important to keep in mind the strong assumptions which are used to present them. For a given PBH model predicting specific distributions, all the constraints have to be reinterpreted using these specific distributions. As shown in Ref. \cite{Arbey:2019vqx}, the distributions can strongly modify the excluded regions.

\begin{figure}[t]
\centering{\includegraphics[height = 7.5cm]{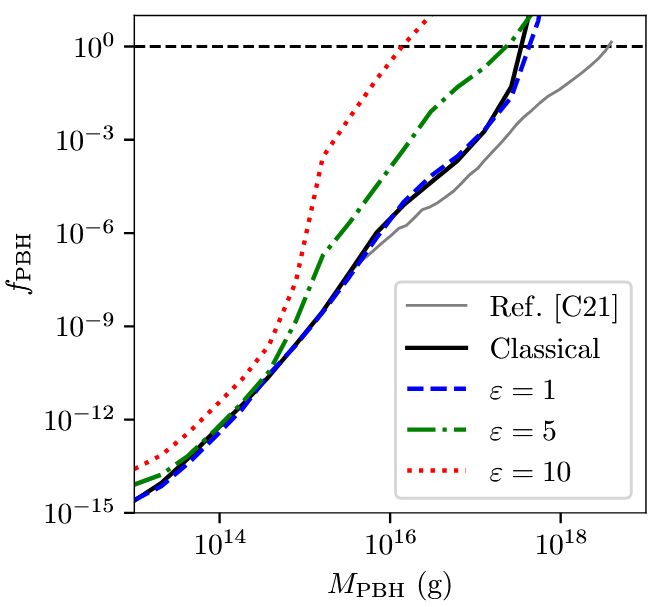}}
\caption{Exclusion limits in the plane fraction of PBHs to dark matter vs. monochromatic PBH mass for different values of $\epsilon$, a parameter of LQG inspired BH metrics. The region on the left is excluded. From Ref. \protect\cite{Arbey:2021yke}.\label{fig:LQG}}
\end{figure}

Another caveat comes from the fact that most of the results assume that the no-hair conjecture holds. However it is clear that PBHs are at the edge of quantum gravity, and that even the metric can be affected by new physics effects. Recent work on the subject reveals that such effects may modify Hawking radiation emission rates \cite{Arbey:2021jif,Arbey:2021yke}, change the PBH lifetime and alter the constraints. This is shown for example in Fig.~\ref{fig:LQG} within a model of Loop Quantum Gravity (LQG) inspired black hole, in which the exclusion limits are shifted when changing the model parameters.

\section{Perspectives}

Primordial black holes are attracting a lot of attention since the direct detection of gravitational waves. In fact, they represent the most general case of black holes in general relativity, and are at the center of different fields of physics. 

They are first gravitational objects which allow to test general relativity. They question the natures of singularities, horizons and spacetime. They may be linked to wormholes, white holes, extra-dimensions, and represent portal to new physics.

Second, primordial black holes are quantum objects, and semi-classical approaches have led to Hawking radiation. BH evaporation is also related to the question of physics at Planck scale and more generally to quantum gravity.

They are also very interesting astrophysical objects, with open questions about the formation mechanisms and the fundamental nature of black holes. From the cosmological and astroparticle points of view, they can represent candidates for dark matter, are related to phase transitions and inflation, and linked to high energy physics, which allow for tests of the phenomena happening in the early Universe.\\
\\
\indent Current researches on primordial black holes span different domains, and they are under scrutiny. For the more formal aspects, they open ways towards new models of black holes, via the quantum gravity questions, but also in the domains of information theory and thermodynamics. Also, in order to interpret the observations, it requires models, simulations and refined numerical relativity techniques, first for simulating the mergers of black holes observed by gravitational wave experiments, but also for understanding black hole formation mechanisms, and the formation and dynamics of large scale structures in presence of black holes \cite{{Boldrini:2019isx}}.

Concerning the observations themselves, since PBHs can have any mass, a very broad spectrum of techniques is needed, in particular cosmological and astrophysical searches with gravitational lensing and telescopes, searches for gravitational waves from mergers, searches for the stochastic gravitational wave background to set constraints on formation mechanisms and Hawking emission, and multi-messenger searches using astroparticle experiments with photons, electrons and positions, neutrinos, antiprotons, etc.

To conclude, primordial black holes are peculiar objects which are of interest for all domains of fundamental physics and philosophy, and they deserve the attention that they currently attract.

\end{document}